\documentclass[twocolumn]{aastex62}

\graphicspath{{./}{figures/}}
\usepackage[utf8]{inputenc}
\usepackage{graphicx}
\usepackage{dcolumn}
\usepackage{bm}
\usepackage{ulem} 
\usepackage{hyperref}

\usepackage{multirow,enumitem}
\usepackage{amsmath}
\usepackage{color}
\usepackage{xcolor}
\usepackage{multirow}

\newcommand{\Om}{\Omega_m}
\newcommand{\LCDM}{\rm{\Lambda}CDM}

\begin{document}

\title{Evidence for Emergent Dark Energy}
\author{Xiaolei Li}
\affiliation{College of Physics, Hebei Normal University, Shijiazhuang 050024, China}
\author{Arman Shafieloo}
\affiliation{Korea Astronomy and Space Science Institute, Daejeon 34055, Korea}
\affiliation{University of Science and Technology, Yuseong-gu 217 Gajeong-ro, Daejeon 34113, Korea}
\date{\today}
 
\begin{abstract}
We introduce a generalised form of an emergent dark energy model with one degree of freedom for the dark energy sector that has the flexibility to include both $\Lambda$CDM model as well as the Phenomenologically Emergent Dark Energy (PEDE) model proposed by~\citet{li2019phenomenologically} as two of its special limits.
 The free parameter for the dark energy sector, namely $\Delta$, has the value of $0$ for the case of the $\Lambda$ and $1$ for the case of PEDE. 
  Fitting the introduced parametric form to Planck CMB data and most recent $H_0$ results from local observations of Cepheids and Supernovae, we show that the $\Delta=0$ associated with the $\Lambda$CDM model would fall out of 4$\sigma$ confidence limits of the derived posterior of the $\Delta$ parameter. Moreover, $H_0$ tensions will be alleviated with emergent dark energy model and this model can satisfy the combination of Planck CMB data and local $H_0$ observations with $\Delta{\rm{DIC}}\,=\,-2.88$ compared with $\LCDM$ model.

\end{abstract}


\section{Introduction}
The standard $\LCDM$ model of cosmology has been greatly successful in describing cosmological observations including type Ia Supernovae (SNe Ia) \citep{riess1998observational,riess2007new,scolnic2018complete}, Baryon Acoustic Oscillation (BAO)\citep{beutler20116df,ross2015clustering,alam2017constraining,zhao2018clustering} and Cosmic Microwave Background (CMB)\citep{ichiki2009brute,komatsu2011seven,ade2016planck,aghanim2018planck}. Despite of its success, comparing the Hubble constant values estimated from local observations of Cepheid in the Large Magellanic (LMC)\citep{riess2019large} and the predicted values from Planck cosmic microwave background (CMB) observations assuming $\LCDM$ model \citep{aghanim2018planck} represent a serious discrepancy. Considering some most recent observations, the tension can reach to about 5$\sigma$ in significance. This implies that either there are considerable, but not accounted systematic-errors in our observations, or, we might need to consider modifications to the standard $\LCDM$ model.

Many ideas have been put forward to resolve the tensions, such as interaction dark energy models \citep{Kumar:2017dnp,DiValentino:2017iww,zheng2017ultra,Yang:2018euj,Yang:2018uae,Kumar:2019wfs,Pan:2019gop,di2019interacting},  metastable dark energy models \citep{shafieloo2017metastable,li2019revisiting}, Quintom dark energy model \citep{panpanich2019resolving} and so on \citep{DiValentino:2015ola,DiValentino:2016hlg,DiValentino:2017zyq,Sola:2017znb,DiValentino:2017oaw,DiValentino:2017rcr,Khosravi:2017hfi,Joudaki:2017zhq,Colgain:2018wgk,DEramo:2018vss,Guo:2018ans,Yang:2018qmz,Poulin:2018cxd,Yang:2019jwn,DiValentino:2019exe,Visinelli:2019qqu,Schoneberg:2019wmt,kreisch2019neutrino,Keeley:2019esp,di2019non,vagnozzi2019new,yang2019dawn,capozziello2019extended}.
In the work of \citet{li2019phenomenologically}, the authors introduced a zero freedom dark energy model -Phenomenologically Emergent Dark Energy (PEDE)-- where dark energy has no effective presence in the past and emerges at the later times.       
All the results for PEDE model show that if there is no substantial systematics in Planck  CMB  data  and  assuming  reliability  of  the  current local $H_0$ measurements, there is a very high probability that with slightly more precise measurement of the Hubble  constant, PEDE model could rule out the cosmological constant with decisive statistical significance.

In this work, we propose a generalised parameterization form for PEDE model that can include both cosmological constant and PEDE model. This generalised parametric form has two parameters to describe the properties of dark energy evolution: one free parameter namely $\Delta$ to describe the evolution slope of dark energy density and parameter $z_t$ that describes the transition redshift where dark energy density equals to matter density. The transition redshift $z_t$ locates where dark energy density equals to matter density and is not a free parameter. This Generalised Emergent Dark Energy (GEDE) model has the flexibility to include both $\LCDM$ model as well as the PEDE model as two of its special limits with $\Delta=0$ and $\Delta=1$, respectively. 
{We emphasis here that such flexibility and generality are particularly important to our research on the behaviour of dark energy evolution, not only because they increase the range of possibilities to be tested but also because they could reduce the possibility for misleading results that an incorrect parameterization form of dark energy evolution could produce.}
{In practice, the parameterization form of dark energy density might be slightly different from PEDE while it still follows emergent dark energy behavior. It can have some interesting implications since dark energy didn't exist in the previous time and it basically appears at late time. }

We confront this model with CMB from Planck 2018 measurements \citep{aghanim2018planck} and most recent $H_0$ results from local observations of Cepheid in the Large Magellanic (LMC) \citep{riess2019large}. We show that the constraints on the parameter $\Delta$ using the combination of local $H_0$ measurement and Planck 2018 CMB results, can rule out the standard $\Lambda$CDM model at more than 4$\sigma$ level where the suggested data combination suggests an emergent behavior for dark energy.

\section{Generalised Emergent Dark Energy (GEDE) model} \label{sec:cos_model}
Assuming a spatially flat universe and the Friedmann-Lema\^{\i}tre-Robertson-Walker (FLRW) metric, the Hubble parameter could be written as
\begin{equation}\label{eq:Ea}
    {H^2(a)}\,=\,H_0^2 \left[\widetilde{\Omega}_{\rm{DE}}(a)+\Om a^{-3} +\Omega_{\rm{R,0}} a^{-4} \right]
\end{equation}
where $a=1/(1+z)$ is the scale factor, $\Om$ and $\Omega_{\rm{R,0}}$ is the current matter density and radiation density, respectively. Here $\widetilde{\Omega}_{\rm{DE}}(a)$ is defined as
\begin{align}
   \widetilde{\Omega}_{\rm{DE}}(a) & \,  =\,\frac{\rho_{\rm{DE}}(a)}{\rho_{\rm {crit,0}}}\,=\, \frac{\rho_{\rm{DE}}(a)}{\rho_{\rm {crit}}(a)}\times \frac{\rho_{\rm {crit}}(a)}{\rho_{\rm {crit,0}}}  \\ 
  & \, =\,{\Omega}_{\rm{DE}}(a)\times \frac{H^2(a)}{H_0^2}
\end{align}
and $\rho_{\rm {crit,0}}\,=\,\frac{3H_0^2}{8\pi G}$, $\rho_{\rm {crit}}(a)\,=\,\frac{3H^2(a)}{8\pi G}$. In $\LCDM$ model, $ \widetilde{\Omega}_{\rm{DE}}(a)\,= (1-\Om-\Omega_{\rm{R,0}})\,=\,$ constant.

\begin{figure}[h!]
\centering
\includegraphics[width=0.45\textwidth]{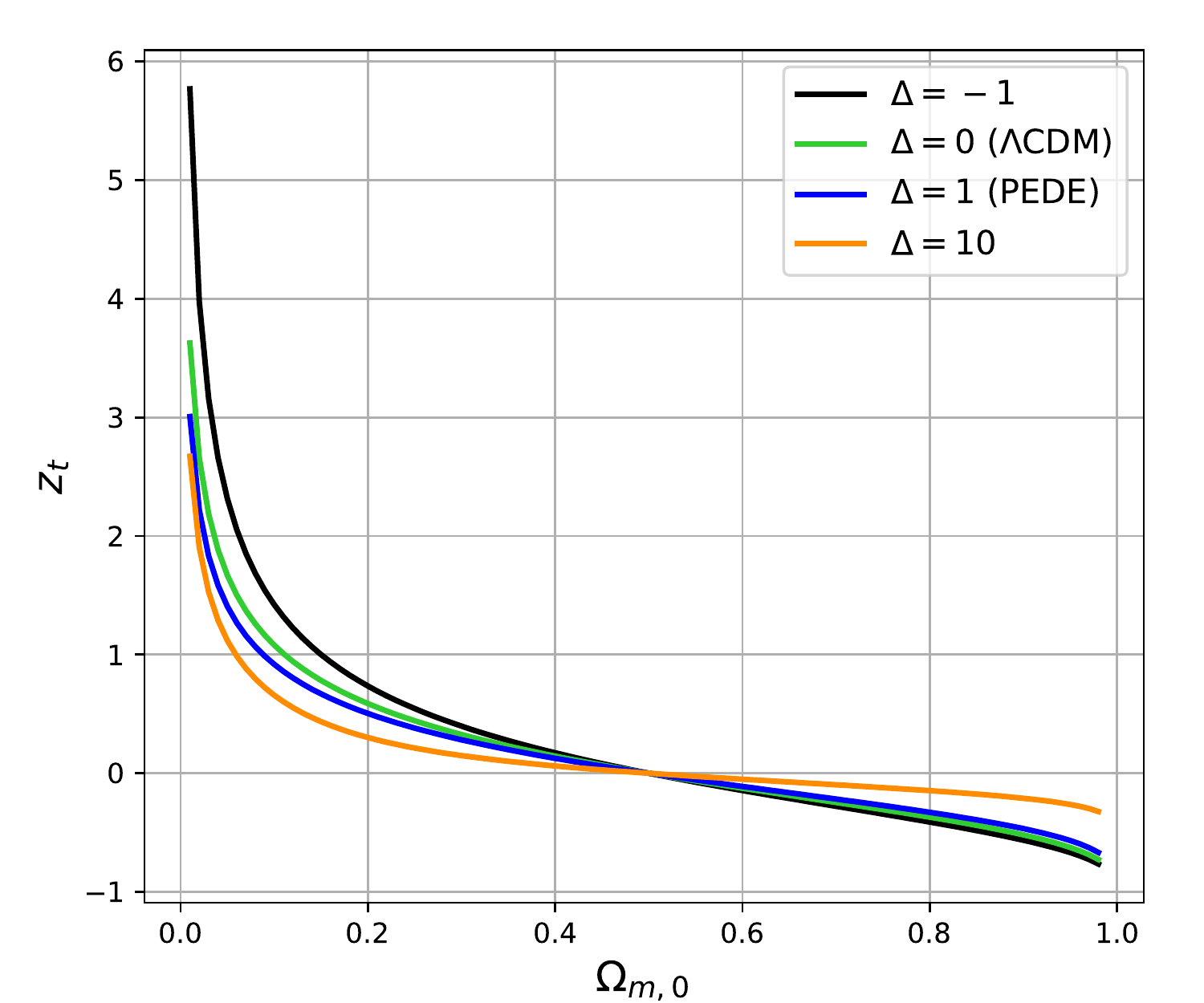}
\caption{Transition redshift $z_t$ as a function of matter density $\Om$ for some certain values of parameter $\Delta$. }
\label{fig:zt_om}
\end{figure}

In GEDE model, the evolution for dark energy density has the following form:
\begin{equation} \label{eq:odez}
{\widetilde{\Omega}_{\rm{DE}}(z)}\,=\, \Omega_{\rm{DE,0}}\frac{ 1 - {\rm{tanh}}\left(\Delta \times {\rm{log}}_{10}(\frac{1+z}{1+z_t}) \right) }{{1+ {\rm{tanh}}\left(\Delta \times {\rm{log}}_{10}({1+z_t}) \right)}}
\end{equation}
here ${\Omega}_{\rm{DE,0}}\,= (1-\Om-\Omega_{\rm{R,0}}) $ and transition redshift $z_t$ can be derived by the condition of ${\widetilde{\Omega}_{\rm{DE}}(z_t)}\,=\,\Om(1+z_t)^3$ (hence it is not a free parameter). In this model, when setting $\Delta\,=\,0$, this model recovers $\LCDM$ model and when setting $\Delta\,=\,1$, it becomes the PEDE model which was introduced in \citet{li2019phenomenologically}, except that in \citet{li2019phenomenologically} the authors set $z_t\,=\,0$  for simplicity while parameter $z_t$ in this work is treated as a transition redshift parameter related to matter density $\Om$ and $\Delta$. 
In Figure~\ref{fig:zt_om}, we show $z_t$ as a function of $\Om$ for some certain values of $\Delta$ for demonstration. 

We can derive the equation of state of GEDE model following:
\begin{equation}
    w(z) \,=\,
    \frac{1}{3} \frac{d\,{\rm{ln}}\, {\widetilde{\Omega}_{\rm{DE}}}}{d z} (1+z)-1
\end{equation}
where we get, 
\begin{equation}
    w(z)\,=\,-\frac{\Delta}{3 {\rm{ln}}\, 10} \times\left({1+{\rm{tanh}}\left(\Delta \times {\rm{log}}_{10}(\frac{1+z}{1+z_t}) \right) }\right) -1. 
\end{equation}

While the derived equation of state of dark energy seems to have a complicated form, it has in fact a simple physical behavior related to dark energy density. We should note that for the two special cases of $\Delta=0$ and $\Delta=1$ we would get the $\Lambda$CDM model and PEDE-CDM model, respectively.

\begin{figure*}
\centering
\includegraphics[width=0.41\textwidth]{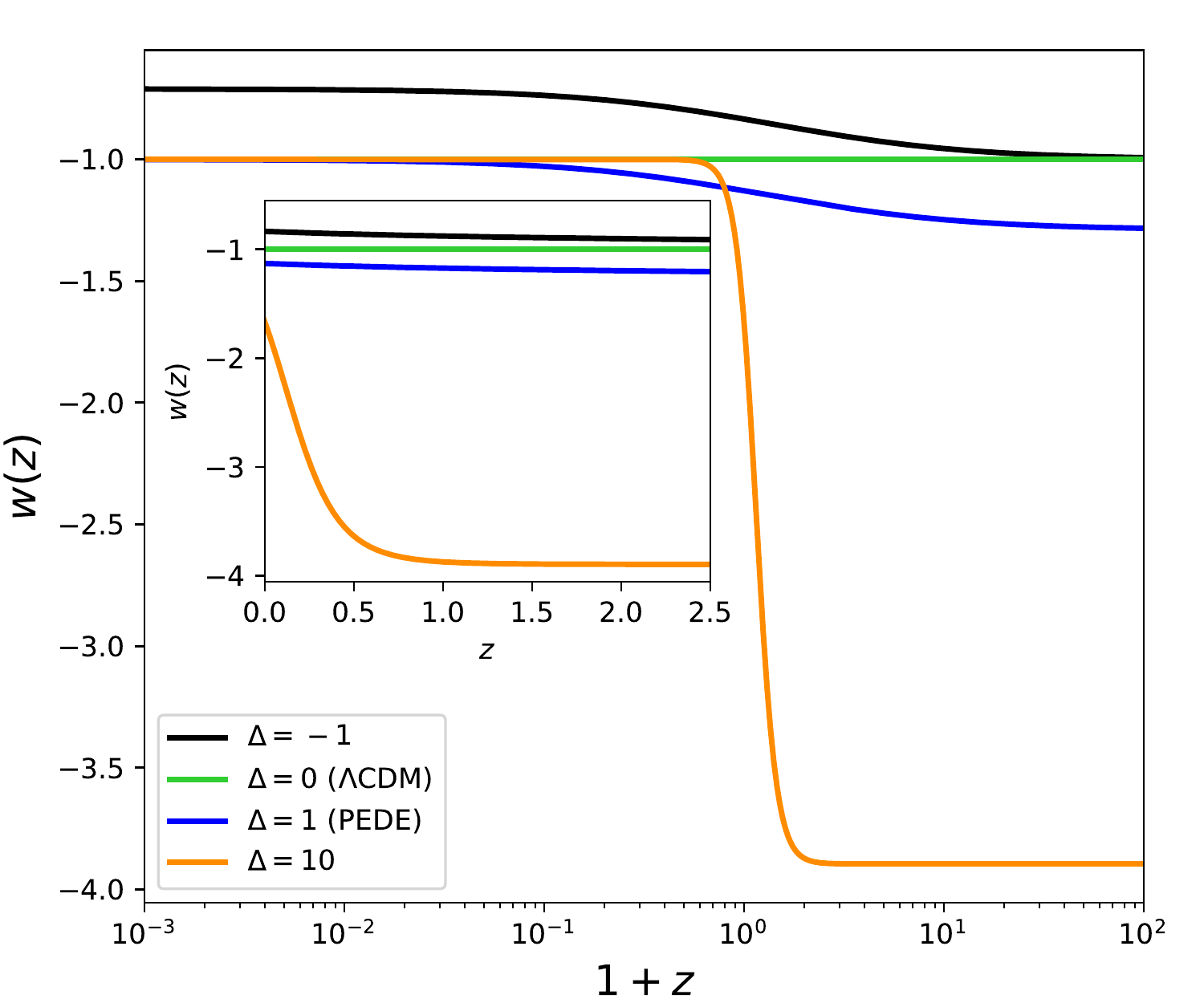}
\includegraphics[width=0.41\textwidth]{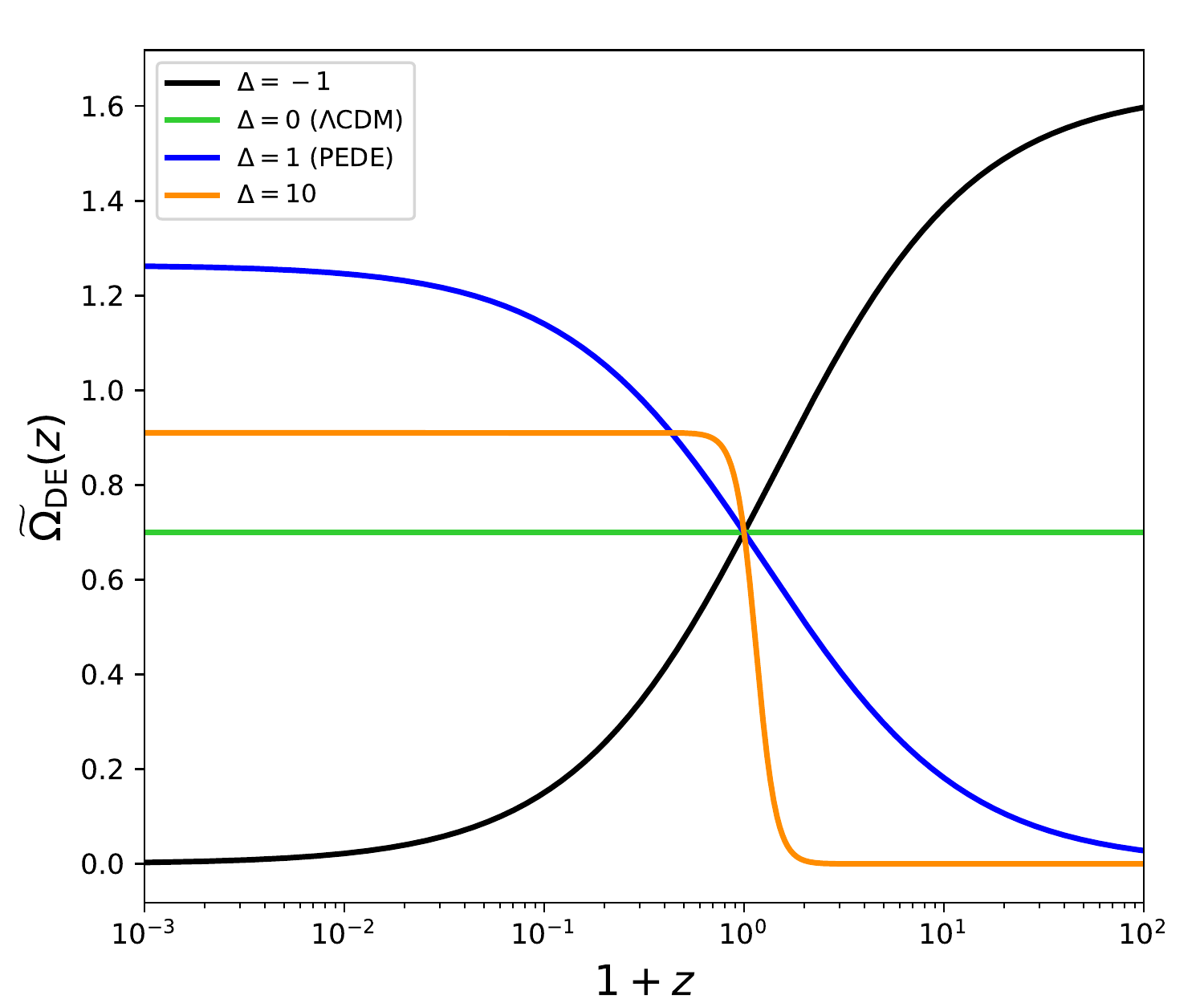}
\includegraphics[width=0.41\textwidth]{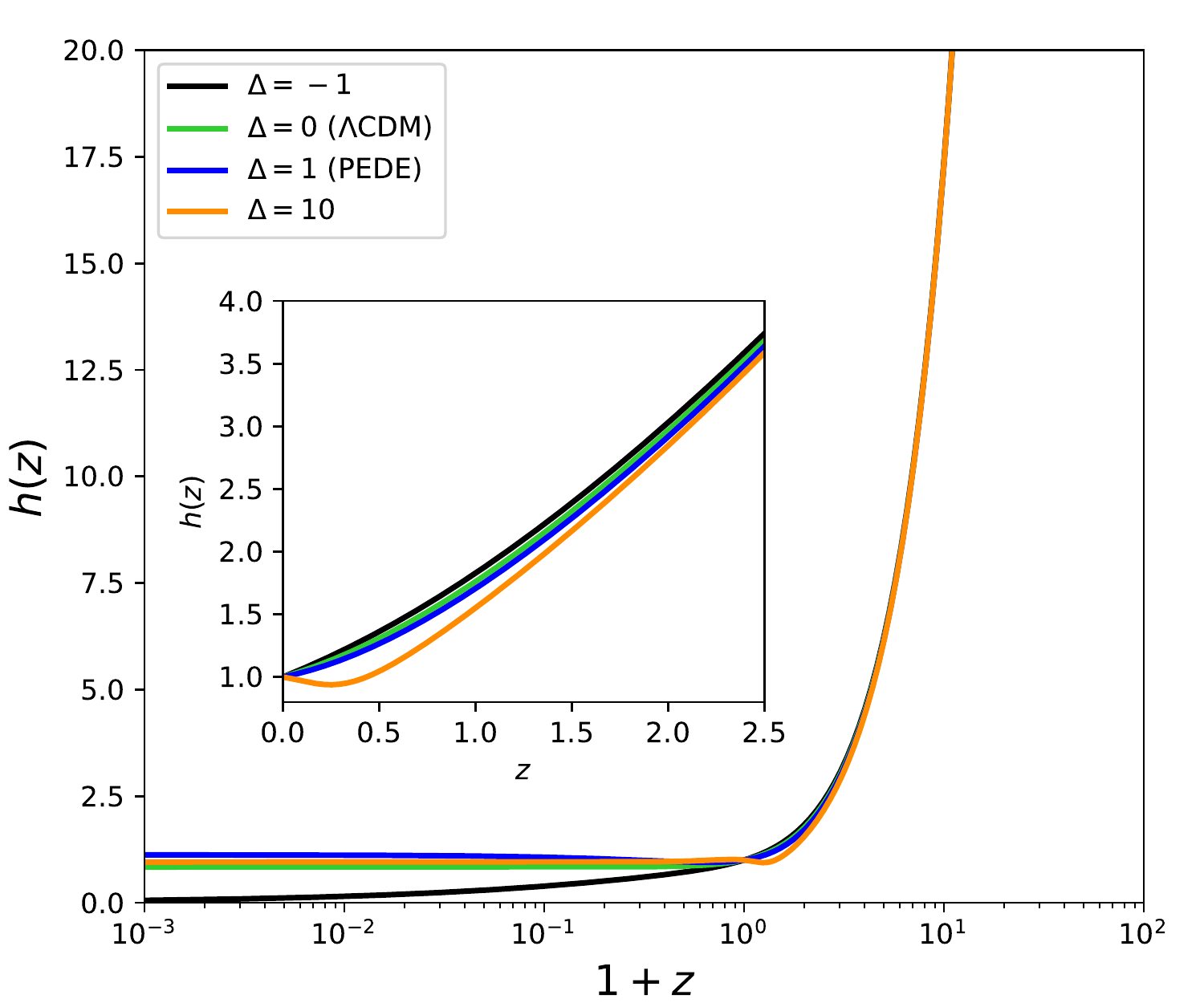}
\includegraphics[width=0.41\textwidth]{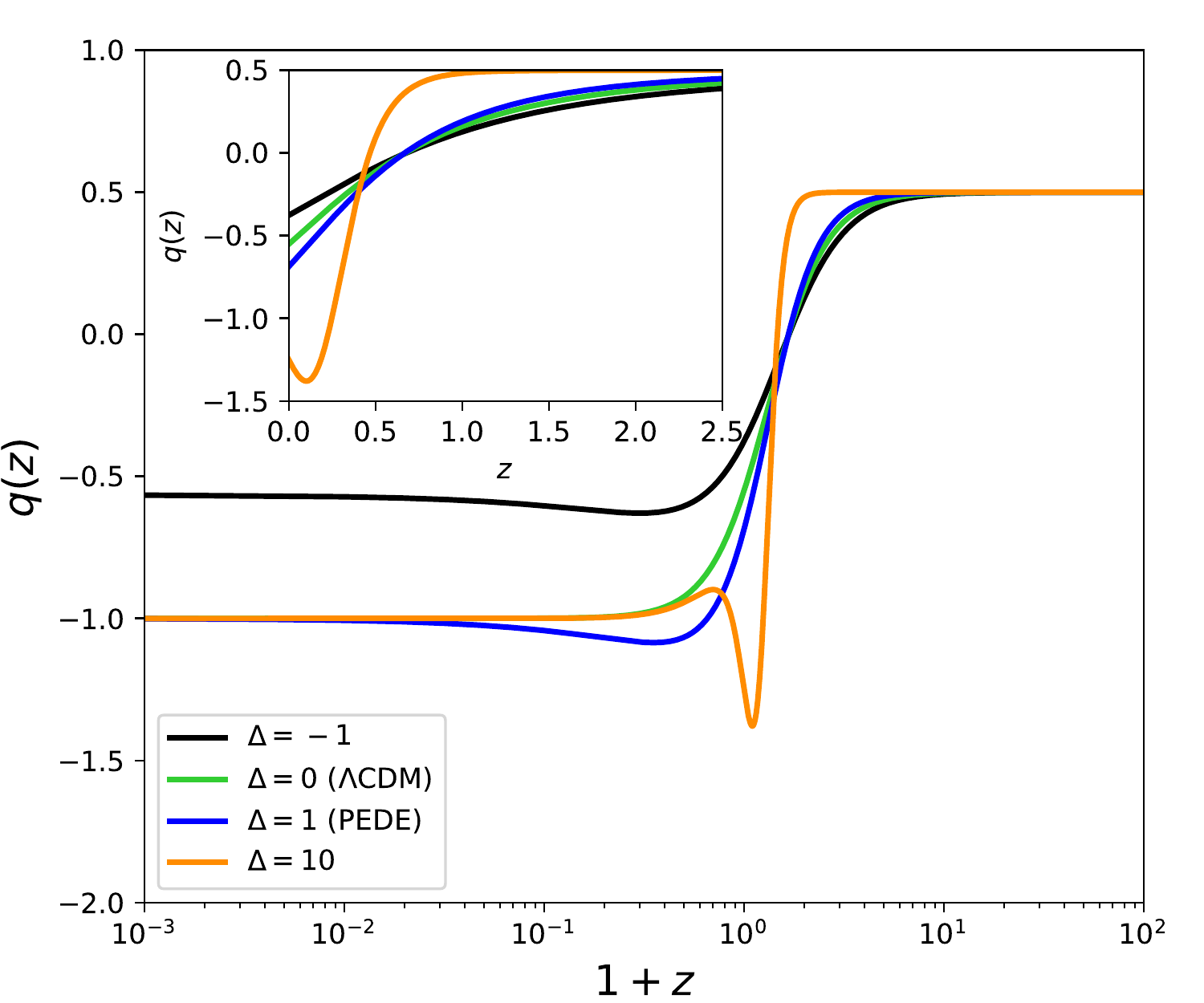}
\caption{Upper left: Equation of state of dark energy $w(z)$ evolved with reshift. Upper right: the evolution of dark energy density ${\widetilde{\Omega}_{\rm{DE}}}(z)$. 
 lower left: Expansion rate $h(z)\,=\,\frac{H(z)}{H_0}$ as a function of redshift $z$. Lower right: deceleration parameter $q(z)$ as a function of redshift $z$. Inner plots show the evolution for each cosmological quantity in linear scale from $z\,=\,0$ to $z\,=\,2.5$.
 $\Om\,=\,0.3$ and flat universe is assumed and plots are mainly for demonstration.}
\label{fig:wz_ode}
\end{figure*}

\begin{table}[]
    \caption{Priors used on various free parameters of $\LCDM$ model, PEDE model and GEDE model during statistical analysis}
    
    \centering
    \begin{tabular}{c|c}
    \hline
    Parameter s& Prior  \\
    \hline
$\Omega_b h^2$          & [0.005, 1]    \\
$\Omega_c h^2$          & [0.001,0.99]  \\
$100\theta_{MC}$        & [0.5,10]  \\
$\tau$                  & [0.01,0.8]  \\
${\rm{ln}}(10^{10} A_s)$& [1.6,3.9]  \\
$n_s$                   & [0.8,1.2]   \\
$\Delta$                & [-1,10]  \\
\hline  
    \end{tabular}
 \label{tab:param_prior}
\end{table}

\begin{table*}

\centering
\caption{We report the 1$\sigma$ constraint results on the free and derived parameters (with *) of $\LCDM$, PEDE model and GEDE model using CMB and CMB+R19. In the last two rows of the table we also display the  $\Delta \chi^2$ and the $\Delta$DIC values with respect to $\LCDM$ model from same data combinations. } 
\label{tab:best_fit}
\begin{tabular}{c|c c|cc | c c}
\hline 
 \multirow{2}{*}{Parameters}  & \multicolumn{2}{c|}{$\LCDM$} & \multicolumn{2}{c|}{{{PEDE}}}   & \multicolumn{2}{c}{GEDE}    \\
  & CMB & CMB+R19  & CMB & CMB+R19 & CMB & CMB+R19 \\
 \hline
$\Omega_b h^2$          &$0.02236\pm 0.00015$  & $0.02255\pm 0.00015 $        & $0.02233\pm 0.00015$ & $0.02239\pm 0.00014$   &$0.02236\pm 0.00015 $ & $0.02236\pm 0.00015        $  \\
$\Omega_c h^2$          &$0.1202\pm 0.0014   $ & $0.1179\pm 0.0013          $ & $0.1204\pm 0.0014  $ & $0.1197\pm 0.0012  $   &$0.1202\pm 0.0014          $ & $0.1201\pm 0.0014  $  \\
$100\theta_{MC}$        & $1.04091\pm 0.00031$ & $1.04121\pm 0.00030        $ & $1.04088\pm 0.00031$ & $1.04096\pm 0.00030$   &$1.04090\pm 0.00031        $ & $1.04092\pm 0.00032        $  \\
$\tau$                  &$0.0543\pm 0.0079   $ & $0.0580^{+0.0074}_{-0.0085}$   & $0.0545\pm 0.0078  $ & $0.0558\pm 0.0079  $   &$0.0542\pm 0.0079          $ & $0.0552\pm 0.0079          $  \\
${\rm{ln}}(10^{10} A_s)$&$3.045\pm 0.016 $     & $3.047\pm 0.017            $ & $3.046\pm 0.016    $ & $3.046\pm 0.016    $   &$3.044\pm 0.016            $ & $3.046\pm 0.016            $  \\
$n_s$                   &$0.9648\pm 0.0044 $   & $0.9704\pm 0.0043          $  & $0.9645\pm 0.0044  $ & $0.9662\pm 0.0041  $   &$0.9647\pm 0.0043          $ & $0.9653\pm 0.0044          $  \\
\hline
$H_0 *$   &  { \multirow{2}{*}{$67.28\pm 0.62 $}}     & { \multirow{2}{*}{$68.35\pm 0.58             $}}  &{ \multirow{2}{*}{$72.24\pm 0.75     $}} & { \multirow{2}{*}{$72.65\pm 0.67     $}}   &{ \multirow{2}{*}{${66.76^{+2.9}_{-0.76} }     $}} & { \multirow{2}{*}{$73.2\pm 1.4               $}}  \\
(\rm{km/s/Mpc}) & & & & & & \\
$\Om *$            &$0.3165\pm 0.0086 $  & $0.3021\pm 0.0076          $  & $0.2748\pm 0.0081  $ & $0.2705\pm 0.0071  $   &$0.323^{+0.029}_{-0.012}   $ & $0.2672^{+0.0097}_{-0.011} $  \\
$r_{\rm drag} H_0 * $ & $9890\pm 100.0$  &   $10080\pm 100$                          &$10620\pm 130      $ & $10690\pm 110 $    & $9820^{+120}_{-430}$ & $10770\pm 210 $ \\
\hline
$z_t *$                 &-     & - &  $0.330\pm 0.016$    & $0.338\pm 0.014   $   &$0.292^{+0.019}_{-0.026} $ & $0.339\pm 0.015$  \\

$\Delta$                & 0  & 0                                                     & 1  & 1                                           &{$< 1.55 $} (3$\sigma$) & $1.13\pm 0.28              $  \\
\hline
$\Delta \chi ^2$ & 0 & 0 & 3.638 & -15.332 & -1.42 &  -15.716 \\
$\Delta$ DIC     & 0 & 0 & 5.29  & -6.02   & 0.6   &  -2.88  \\
\hline
\end{tabular}
\end{table*}

\begin{figure*}
\centering
\includegraphics[width=0.35\textwidth]{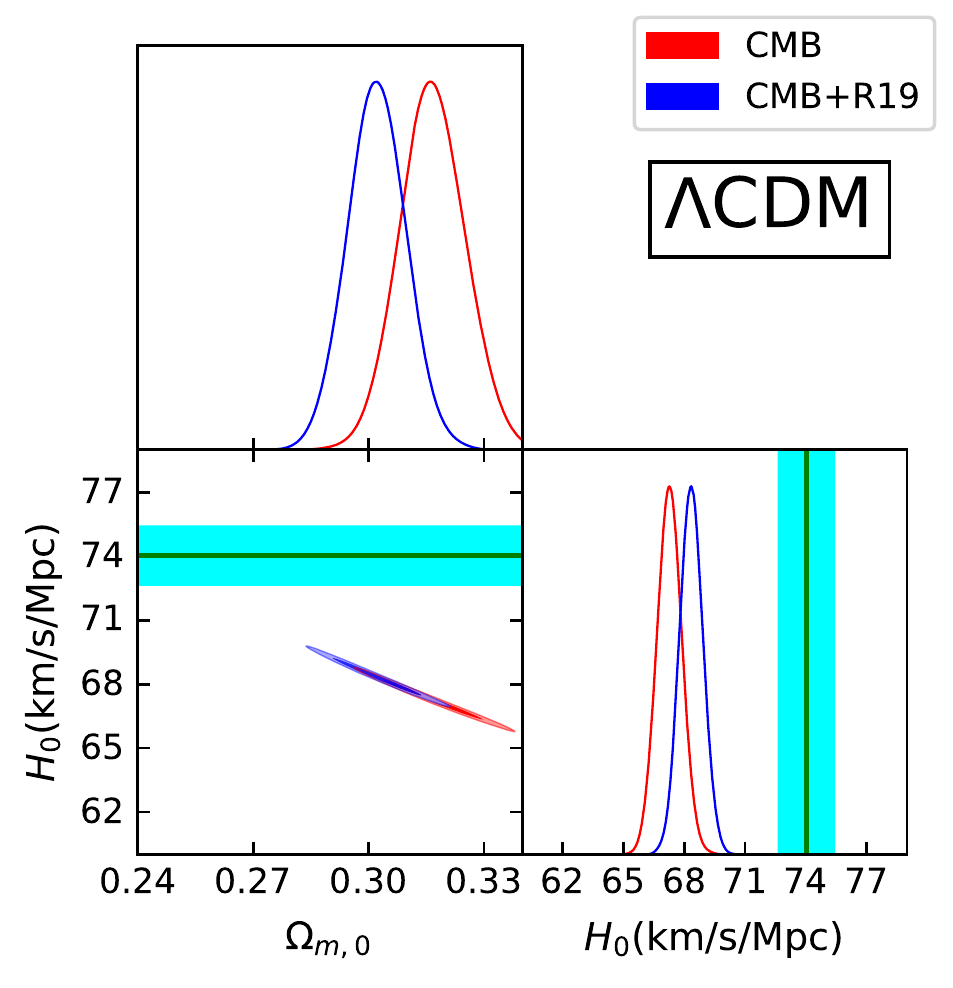}
\includegraphics[width=0.62\textwidth]{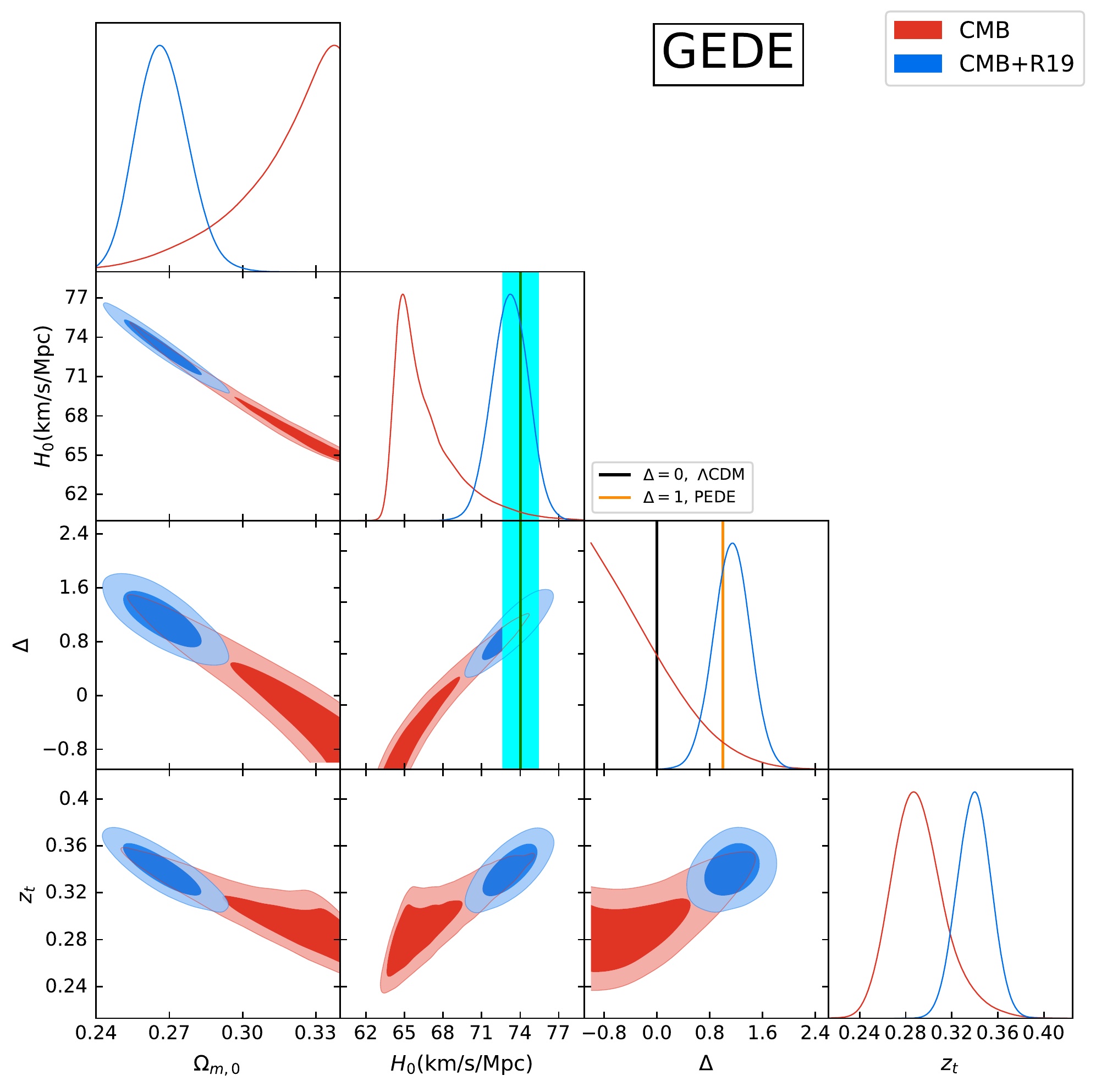}
\caption{Constrain results for $\LCDM$ model (left plot) and GEDE model (right plot) from CMB and CMB+R19. The cyan shadows show the 1$\sigma$ $H_0$ results from \citet{riess2019large}. In the 1D likelihood of GEDE for $\Delta$, we show $\Delta\,=\,0$ for $\LCDM$ model in black vertical line and $\Delta\,=\,1$ for PEDE model in dark orange vertical line. Note that $z_t$ is not a free parameter and is shown for clarity. }
\label{fig:lcdm_omh0}
\end{figure*}

Assuming a spatially flat universe and $\Om \,=\,0.3$, we show properties for GEDE model as a function of redshift for some certain values of $\Delta$ in Figure~\ref{fig:wz_ode} for demonstration. Upper left plot shows the evolution of equation of state $w(z)$ while upper right plot shows the evolution of dark energy density ${\widetilde{\Omega}_{\rm{DE}}(z)}$ from redshift $10^{-3}$ to $10^2$. Lower plots show the evolution of expansion rate $h(z)\,=\,H(z)/H_0$ (left) and deceleration parameter $q(z)$ (right). Different colors correspond to parameter $\Delta$ fixed at some certain values for demonstration. In terms of linear scale of redshift from 0 to 2.5, we show the evolution for $w(z)$, $h(z)$ and $q(z)$ in each inner plot, respectively.\\

In our analysis, we consider CMB measurement from Planck TT, TE, EE+lowE data released in 2018 \citep{aghanim2018planck}. In addition to CMB measurements, we add local measurement $H_0\,=\, 74.03\pm 1.42 $ from \citet{riess2019large} in our analysis (denote as R19 hereafter). The constraint results are obtained with Markov Chain Monte Carlo (MCMC) estimation using \texttt{CosmoMC} \citep{lewis2002cosmological} \textbf{at the background level}. Parameter priors for $\LCDM$ model, PEDE model and GEDE model during statistical analysis are showed in Table~\ref{tab:param_prior}

 For quantitative comparison between GEDE model, PEDE model and $\Lambda$CDM model, we employ the deviance information criterion (DIC) \citep{spiegelhalter2002bayesian,liddle2007information}, defined as 
\begin{equation}
   { \rm{DIC}}\,\equiv \, D(\Bar{\theta})+2p_D\,=\,\overline{D(\theta)}+p_D,
\end{equation}\label{eq:DIC}
where $ p_D\,=\,\overline{D(\theta)}-D(\Bar{\theta})$ and $D(\Bar{\theta})\,=\,-2\, {\rm{ln}}\, \mathcal{L}+C$, here $C$ is a 'standardizing' constant depending only on the data which will vanish from any derived quantity and $D$ is the deviance of the likelihood.

{The value of DIC for a single model is meaningless in this kind of study. What is useful is the difference in values of DIC between cosmological models, $\Delta$DIC. Therefore in our analysis, we calculate the values of $\Delta$DIC with respect to $\LCDM$ model for same observations. Negative values means the model fits the observations better than $\LCDM$ model  while positive values means the opposite.}
\\

We will show that by adding local measurement $H_0\,=\, 74.03\pm 1.42 $ from \citet{riess2019large} to CMB measurements, GEDE model behaves better than $\LCDM$ model and in the context of the GEDE parametric model $\Lambda$CDM model stays outside of the 4$\sigma$ confidence limit.

\section{Results and discussion} \label{sec:res}
\begin{figure*}
\centering
\includegraphics[width=0.41\textwidth]{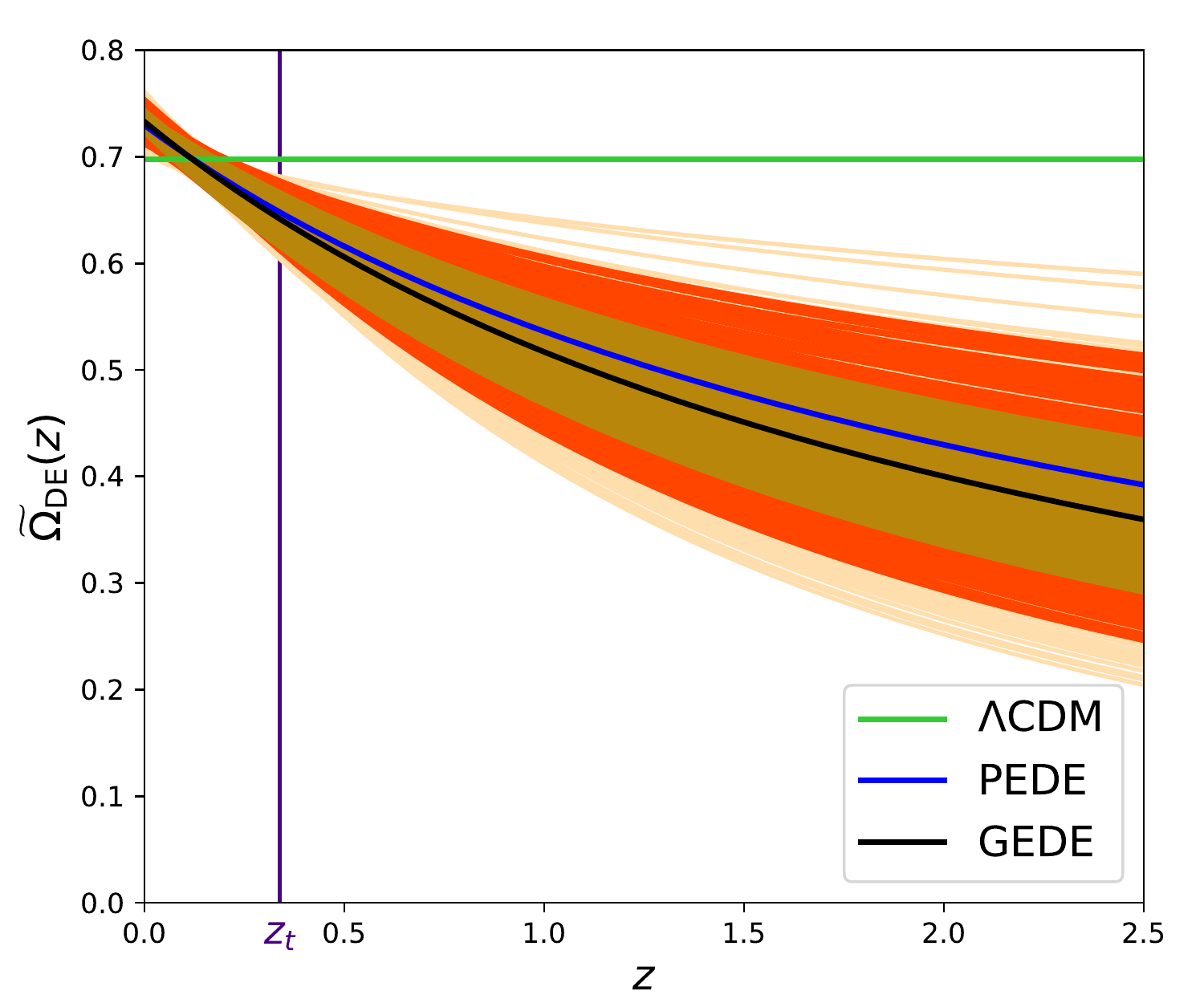}
\includegraphics[width=0.41\textwidth]{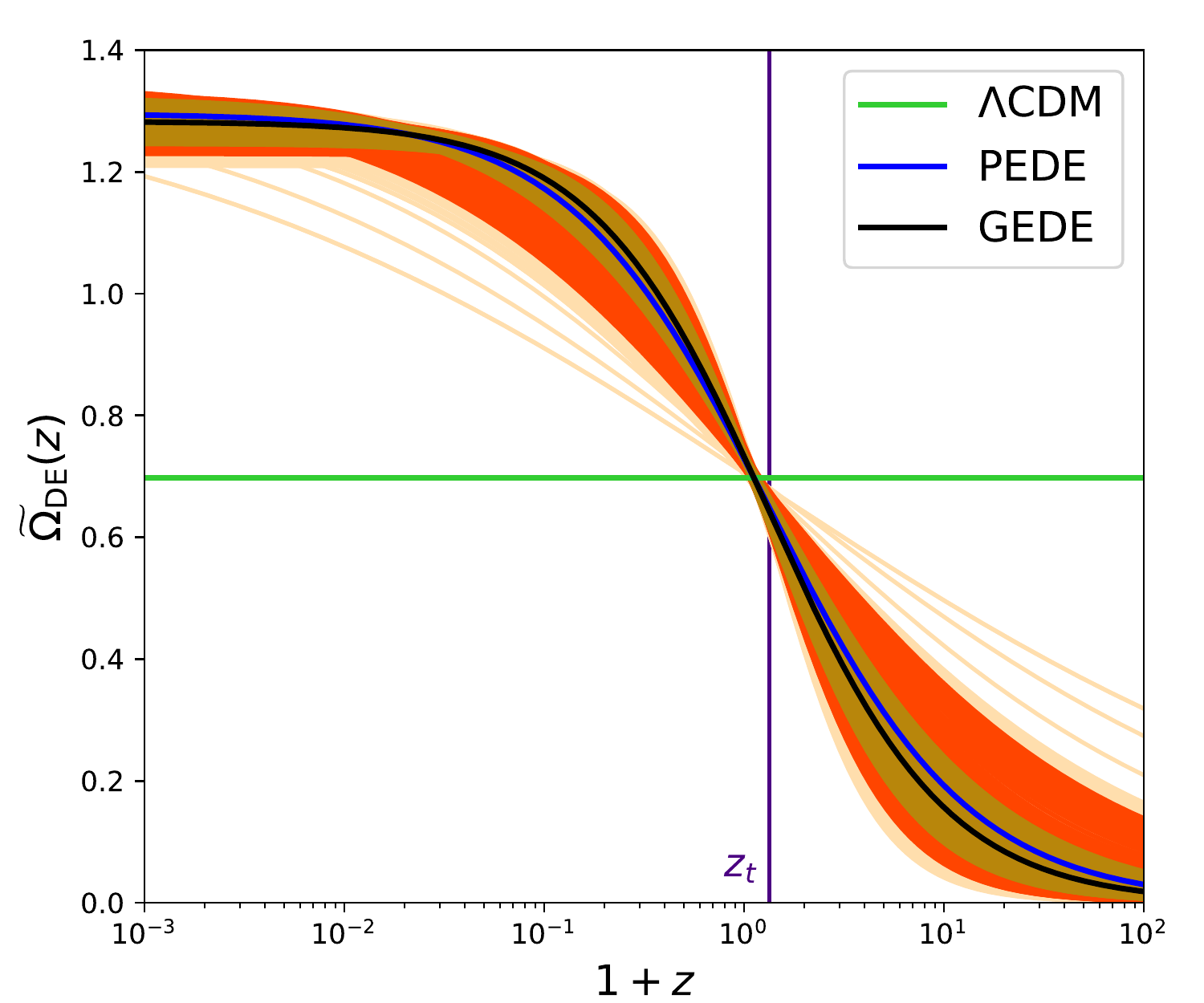}
\caption{The evolution of dark energy density ${\widetilde{\Omega}_{\rm{DE}}}(z)$. Left plot show the evolution in linear scales of $z$ from $0$ to $2.5$ while right plot show the evolution in logarithm scales of $1+z$ from $0$ to $2.5$. The golden, dark orange and the light orange lines are from 1$\sigma$, 2$\sigma$ and 3$\sigma$ confidence range of the GEDE model fitting CMB+R19 data. The black, blue and green solid lines are the best fit results from GEDE, PEDE and $\Lambda$CDM model respectively. It is clear that the CMB+R19 data suggest an emergent dark energy behaviour while cosmological constant is very much outside of the confidence limits. The vertical lines display the mean values of $z_t$ from CMB+R19.}
\label{fig:odez_bf}
\end{figure*}

We summarize the best fit and the 1$\sigma$ constrain results using CMB and CMB+R19 for $\LCDM$ model, PEDE model and GEDE model in Table~\ref{tab:best_fit}. The three cosmological parameter denote with *, {$H_0$, $\Om$ and $r_{\rm{drag}}H_0$}, are derived parameters. Parameter $\Delta$ is fixed to 0 for $\LCDM$ model and 1 for PEDE model and set as free parameter for GEDE model.
In the last two rows we also show $\Delta \chi^2$ and the $\Delta$DIC values with respect to $\LCDM$ model from same data combinations.
{We can see that, both PEDE and GEDE works better with respect to $\LCDM$ model with $\Delta \chi^2 = -15.332$ and $\Delta \chi^2= -15.716$ when using CMB+R19, respectively. When calculating DIC values, as can be seen from Table~\ref{tab:best_fit}, PEDE works the best within the three models for CMB+R19 with $\Delta  \rm{DIC} = -6.02$ with respect to $\LCDM$ model and GEDE works better than $\LCDM$ model with $\Delta \rm{DIC} = -2.88$ for CMB+R19. } 

In Figure~\ref{fig:lcdm_omh0}  we present 1$\sigma$ and 2$\sigma$ contours from CMB and CMB+R19 for $\LCDM$ model (left) and GEDE model (right).

From Table~\ref{tab:best_fit} and Figure~\ref{fig:lcdm_omh0}, it is obvious that with GEDE model, the constraints on $H_0$ from CMB and CMB+R19 is in agreement with each other {at about 1.5$\sigma$ confidence level} and also agree with the local $H_0$ results from \citet{riess2019large}. {One should note the asymmetric probability distribution for Hubble constant as it shows a longer tail at higher values in the 1D posterior result considering CMB data alone.}
  With CMB measurement alone the constraints on $\Delta$ parameter of the GEDE model do not distinguish between $\LCDM$ model and PEDE-CDM model. However, for the case of the combined CMB+R19 data, we can derive $\Delta\,=\,1.13\pm 0.28$ and it is very clear that $\LCDM$ model ($\Delta\,=\,0$) is now outside 4$\sigma$ confidence level region. 

  For comparison between different models, we show the evolution of dark energy density ${\widetilde{\Omega}_{\rm{DE}}}$ as a function of redshift $z$ in Figure~\ref{fig:odez_bf}. The golden, dark orange and the light orange lines corresponds to 1$\sigma$, 2$\sigma$ and 3$\sigma$ confidence range of the dark energy density from GEDE model from its converged MCMC chains fitting CMB+R19 data. The black, blue and green solid lines are the best fit results for GEDE, PEDE and $\Lambda$CDM models respectively. 
  
  From Figure~\ref{fig:odez_bf} it is clear that the CMB+R19 data suggest an emergent dark energy behaviour while cosmological constant is well outside of the 3$\sigma$ confidence limits (it is in fact outside of the 4$\sigma$ confidence limits which are not shown). We should note that having an emergent form of dark energy is not affecting much on the posteriors of the parameters related to early universe such as spectral index of the primordial fluctuations but there are substantial differences calculating the ISW effect for  in comparison with the case of cosmological constant. However, due to cosmic variance, the CMB angular power spectrum data is not much sensitive to these low multiple differences~\citep{pan2019reconciling}. Some more extensive analysis of the GEDE model would be useful to have a better understanding of the behaviour of this model. This will be done in future works.

\section{Conclusion} \label{sec:con}
In this work, Phenomenologically Emergent Dark Energy model (PEDE) that was introduced in \citet{li2019phenomenologically} is generalised to GEDE model (Generalised Emergent Dark Energy) which has one degree of freedom for the dark energy sector and has the flexibility to include both PEDE model and $\LCDM$ model as two of its special limits. We confront this model with CMB measurements from  \texttt{Planck} 2018 \citep{aghanim2018planck} and $H_0$ result from local observations of Cepheid in the LMC \citep{riess2019large}, as two most important and independent cosmological observations at high and low redshifts, and compare the results with the case of $\LCDM$ model using DIC analysis and find that, 
1) our results are consistent with the previous analysis by \citet{li2019phenomenologically}, \citet{pan2019reconciling} and \citet{arendse2019cosmic} that emergent dark energy model works better than $\LCDM$ if we trust CMB measurements as well as local $H_0$ measurements (CMB+R19) and this model can alleviate $H_0$ tension that is present in $\LCDM$ model. 
2) using CMB+R19 data and within the context of the GEDE parameterization, $\LCDM$ model ($\Delta\,=\,0$) is ruled out at 4$\sigma$ confidence level, which is a strong evidence for emergent dark energy behaviour.

We should note that future CMB measurements such as Advanced CMB Stage 4 \citep{abazajian2016cmb} can surpass \texttt{Planck}  CMB measurements in their ability to put tight constraints on cosmological parameters, including Hubble constant $H_0$ assuming any particular cosmological model. Furthermore, local $H_0$ measurement will be also improved with highly improved distance calibration from Gaia \citep{prusti2016gaia} and improved techniques such as using the tip of the red giant branch to build the distance ladder \citep{freedman2017cosmology} as well as using strong lens systems to measure the expansion rate \citep{Suyu:2016qxx,Birrer:2018vtm,Liao:2019qoc}.
All of these improvement will lead to higher precision of $H_0$ measurements and would finally shed light on the nature of the current tensions.\\

{We should note that during the revision stage of this work, some relevant analysis on PEDE as well as the current GEDE model have been appeared in the literature \citet{yang2020complete,10.1093/mnras/staa2052} that can provide more information for the readers about our emergent dark energy model. }

\acknowledgments
 We would like to thank Zong-Hong Zhu and Beijing Normal University for the hospitality during the early stages of this work. We thank the "6th Korea-Japan workshop on dark energy at KMI" at Nagoya University for providing a discussion venue for the co-authors.
 X. Li thanks Weiqiang Yang for useful technical discussions.
X. Li is supported by National Natural Science Foundation of China under Grants No. 11947091, Hebei NSF under Grant No. A202005002 and the fund of Hebei Normal University under Grants No. L2020B20. A.S. would like to acknowledge the support of the Korea Institute for Advanced Study (KIAS) grant funded by the Korea government.
This work benefits from the high performance computing clusters at College of Physics, Hebei Normal University and the high performance computing clusters Seondeok at the Korea Astronomy and Space Science Institute.

\bibliography{references}
\end{document}